\documentclass[11pt]{article}
\usepackage{amssymb,latexsym,amsmath}
\usepackage{xcolor}
\usepackage{bm}
\usepackage{amsmath,amscd}
\usepackage{graphicx}
\usepackage{newtxtext}
\usepackage{newtxmath}
\usepackage{geometry}
\usepackage{hyperref}
\hypersetup{
    colorlinks = true,
    urlcolor   = blue,
    citecolor  = black,
}

\newcommand{\RomanNumeralCaps}[1]
\linenumbers

\title{A family of variational principles of minima for the plasticity, the friction contact and the fracture mechanics}

\author{\textbf{G. de Saxc\'e} \\
Univ. Lille, CNRS, Centrale Lille, UMR 9013 – LaMcube \\
Laboratoire de m\'ecanique multiphysique multi\'echelle, \\
F-59000, Lille, France, Email: gery.de-saxce@univ-lille.fr}

\begin{document}
\maketitle

\begin{abstract}
The paper is a synthesis of several works on the variational principles for application to the mechanics and the physics, inspired from original ideas of Brezis, Ekeland and Nayroles. On this basis, we developed an unified framework for dynamic dissipative systems that leads to a space-time variational principle of minimum constructed  with tools of convex analysis and symplectic geometry. We stress the essential ideas and concepts. They are illustrated with various theoretical and numerical examples. 
\end{abstract}


\vspace{0.3cm}


% (Primary)
% 49J40 Variational methods including variational inequalities 

% (Secondary)
% 53Z05 Application to physics (dans 53Dxx Symplectic geometry, contact geometry)
% 70F40 Problems with friction
% 74C05 Small-strain, rate-independent theories (including rigid-plastic and elasto-plastic materials) 
% 74F10 Brittle fracture

%%%%%%%%%%%%%%%%%%%%%%%%%%%%%%%%%%%%%%%%%%%%%%%%%%%%%%%%%%%%%%%%%%%%%%

\section{Introduction}

The variational principles have first of all a mnemonic value which allows deducing the physical laws in a consistent and systematic way. Nowadays, it is also a powerful tool to construct approximations, especially for numerical simulations. In this paper, we present a class of non incremental space-time variational principles of minima for dissipative dynamic systems with a broad spectrum of applications. 

For dissipative systems, several authors proposed unified frameworks inspired from the Riemannian and symplectic geometries: the {\it metriplectic systems} of Morrison \cite{Morrison 1986}, developed further by Grmela and \"Ottinger as the {\it GENERIC systems} \cite{Grmela 1997}, based on the Hamiltonian and the Onsager entropy, and the {\it rate-independent systems} of Mielke and Theil \cite{mielketh99}, with an energy function and a 1-homogeneous dissipation potential. Inspired from these two approaches and the formalism of Hamiltonian inclusions \cite{bham}, we proposed in \cite{SBEN} a symplectic version (SBEN) of the Brezis-Ekeland-Nayroles principle (BEN). We are working in the phase space of the Physicist of which the elements $\bm{z}$ are composed of the degrees of freedom —which may be also fields— and the associated dynamical momenta. It is equipped with the canonical symplectic form. The key idea is to decompose additively the time rate $\dot{\bm{z}}$ into a reversible part $\dot{\bm{z}}_R$ (the symplectic gradient) and a dissipative or irreversible one $\dot{\bm{z}}_I$, and then to define the symplectic subdifferential $\partial^{\omega} \phi (\bm{z})$ of the dissipation potential. To release the restrictive hypothesis of $1$-homogeneity (in particular to address viscoplasticity), we introduce the symplectic Fenchel polar $\phi^{*\omega}$, that allows to build theoretical methods to model and analyze dynamical dissipative systems in a consistent geometrical framework with the numerical approaches not very far in the background. Numerical simulations with the BEN principle were performed for elastoplastic structures in statics (\cite{Cao 2020}, \cite{Cao 2021a}) and in dynamics \cite{Cao 2021b}.

This work is a survey of various works and topics which share a common variational approach inspired of the convex analysis and the symplectic geometry. We stress the essential ideas and concepts. The technical details are a few and the reader is referred to the bibliography to know more. The paper is structured as follows. Section \ref{Section - A smidgen of convex analysis} is a very short reminder of the definitions and results of convex analysis useful in the sequel, focussing on the Fenchel inequality and the extremality condition. In Section \ref{Section - Brezis-Ekeland-Nayroles principle for standard plasticity and viscoplasticity}, we start with a first simple application of the Brezis-Ekeland-Nayroles (BEN) variational principle, the standard plasticity and viscoplasticity and we discuss particular cases. The Section \ref{Section BEN principle for Signorini unilateral contact} is devoted to a second application, the Signorini unilateral contact without friction between elastic bodies and we show the link with the classical formulation in the form of a  Linear Complementary Problem. Next, we present a generalization framework for dynamical dissipative systems in Section \ref{Section - Unified frameworks for dynamic dissipative systems}, introducing the symplectic subdifferential and the symplectic polar function of a convex function. In Section \ref{Section - A symplectic minimum variational principle for dynamical dissipative media}, we construct a functional from a dissipation potential, the Hamiltonian for the reversible behaviour and the symplectic form for the dynamics and state the symplectic Brezis-Ekeland-Nayroles (SBEN) principle. In Section \ref{Section - Application : standard plasticity and viscoplasticity in dynamics}, we show how to recover the case of the plasticity and viscoplasticity from the previous general formulation. Section \ref{Section - Other applications} is a quick survey of several extensions, the plasticity in finite strain in the Lagrangian specification, and Navier-Stokes and Bingham fluids in the Eulerian specification. In Section \ref{Section - Bipotentials}, we recall the concept of bipotential, a mathematical tool to model the class of non associated constitutive laws as the unilateral contact with Coulomb friction law. In Section \ref{Section - Symplectic bipotentials and SBEN principle for non associated constitutive laws}, we introduce the concept of symplectic bipotential to extend the SBEN principle to this class of constitutive laws. The aim of the Section \ref{Section - Application to the crack propagation with friction between the crack sides} is to illustrate this approach by applying it to the crack propagation with friction between the crack sides. For the brittle fracture, we use a stability domain with the crack extension governed by the normality law. We show the relevance of the normality law in this context by comparison with experimental data.

\section{A smidgen of convex analysis}
\label{Section - A smidgen of convex analysis}

\begin{figure}[ht!]
\centering
\includegraphics[scale=.50]{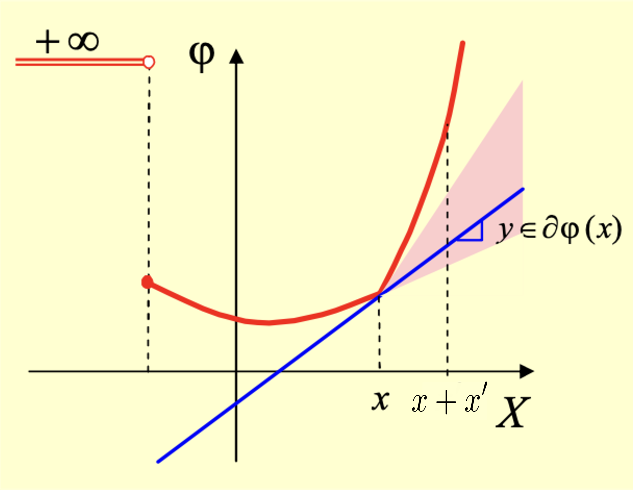}
\caption{Subdifferential of a convex function}
\label{fig Subdifferential of a convex function}
\end{figure}

Our first mathematical tool is the  convex analysis, a modern formalism for nonsmooth constitutive laws (friction, plasticity, viscoplasticity, crack growth and damage). We are working with two spaces $X$ and $Y$ in duality thanks to a dual pairing 
$$ X \times Y \rightarrow \mathbb{R}: (\bm{x} , \bm{y}) \mapsto \langle \bm{x} , \bm{y} \rangle
$$
The \textbf{Fenchel transform} $\varphi^* (\bm{y})$ of a convex function with real and eventually infinite values
$$\varphi: X \rightarrow\mathbb{R} \cup \lbrace + \infty \rbrace : \bm{x} \mapsto\varphi (\bm{x}) 
$$ 
is the \textbf{polar function} 
$$ \varphi^* (\bm{y}) = \sup_{\bm{x}} \; (\langle \bm{x} , \bm{y} \rangle - \varphi (\bm{x}) )
$$
By construction, because of the supremum, it is convex and verifies the \textbf{Fenchel inequality}
$$ \forall \bm{x}',\bm{y}', \quad
\varphi (\bm{x}') + \varphi^* (\bm{y}') - \langle \bm{x}',\bm{y}' \rangle \geq 0  
$$
The \textbf{subdifferential} of a convex function at $\bm{x}$ is a set of generalized gradients, called subgradients, defined by
$$ \partial \varphi (\bm{x}) = \left\lbrace
                            \bm{y} \in Y \quad\mbox{such that} \quad \forall \bm{x}',\qquad 
                            \varphi (\bm{x} + \bm{x}') - \varphi (\bm{x}) \geq \langle \bm{x}' , \bm{y} \rangle
                            \right\rbrace
$$ 
As represented in Figure \ref{fig Subdifferential of a convex function}, a subgradient is the slope of an hyperplan in contact with the graph of $\varphi$ at $\bm{x}$ and anywhere below it. At a regular point of the graph where the function is differentiable, the gradient is the single subgradient. The subdifferential is convex and possibly an empty set.

For a convex and lower semicontinuous function $\varphi$, there is equivalence between the three following conditions:
\begin{center}
\begin{tabular}{l l l l l}
 normality law & & inverse law & &  extremality condition\\
$ \bm{y} \in \partial \varphi (\bm{x})$  & $\Leftrightarrow$ & 
 $\bm{x} \in \partial \varphi^* (\bm{y})$  & $\Leftrightarrow$ &
 $\varphi (\bm{x}) + \varphi^* (\bm{y} ) = \langle \bm{x} , \bm{y} \rangle$ \\
 \end{tabular}
\end{center}
When applying the convex analysis to the nonsmooth mechanics, the most usual way to represent the constitutive law is to use the normality law (or the inverse law). The point of view adopted here is to prefer to represent  it by the extremality condition which will play in this paper a central role together with the Fenchel inequality.

\begin{figure}[ht!]
\centering
\includegraphics[scale=.50]{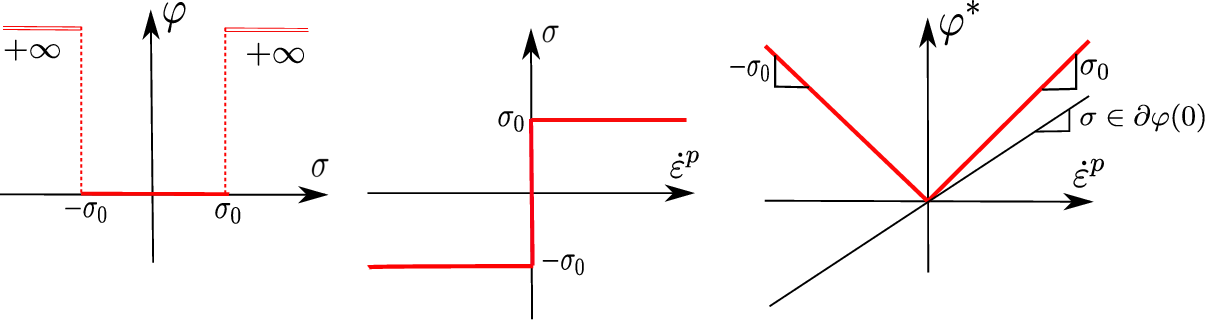}
\caption{Example: 1D associated plasticity}
\label{fig Example: associated plasticity}
\end{figure}

As example of application, we consider the 1-dimensional associated plasticity with a yield stress $\sigma_0$, represented in Figure \ref{fig Example: associated plasticity}:
\begin{itemize}
\item On the center, the constitutive law is the graph of a set-valued function.  If the plastic strain rate $\dot{\varepsilon}^p$ is zero, the stress $\sigma$ belongs to the plastic domain $K = \lbrack - \sigma_0, \sigma_0 \rbrack$. For the plastic yielding in traction: $\dot{\varepsilon}^p > 0$ (resp. in compression:  $\dot{\varepsilon}^p < 0$), the stress is equal to $\sigma_0$ (resp. $ - \sigma_0$). 
\item On the right, we represented the conic graph of the dissipation potential $\varphi^*$. We verify geometrically that $\partial \varphi^* (0) = K$, $\partial \varphi^* (\dot{\varepsilon}^p) = \lbrace \sigma_0 \rbrace$ in traction and $\partial \varphi^* (\dot{\varepsilon}^p) = \lbrace  - \sigma_0 \rbrace$ in compression.
\item  On the left, the graph of the $\varphi$ is a well potential with infinite walls at $\sigma = \pm \sigma_0$, of null value on the plastic domain $K$. This kind of function is called the \textbf{indicator function} of $K$ and will be denoted $I_K$ in the sequel. The physical meaning of an infinite value at a given stress $\sigma$ is that reaching $\sigma$ would require an infinite power, \textit{i.e.} $\sigma$ is not a physical stress state. We verify that $\partial \varphi (\sigma) = \lbrace 0 \rbrace$ if $\sigma \in K$, $\partial \varphi (\sigma_0) = \lbrack 0, +\infty \lbrack $,  $\partial \varphi (- \sigma_0) = \rbrack  -\infty, 0 \rbrack$ and otherwise $\partial \varphi (\sigma) = \emptyset$.
\end{itemize}

Many three dimensional extensions were proposed in the literature (von Mises for metals, Mohr-Coulomb, Drucker-Prager for soils and so on). The viscoplasticity model is similar with a plastic domain $K$ but the creep flow occurs also for stress values larger than the yield stress.

\section{Brezis-Ekeland-Nayroles principle for standard plasticity and viscoplasticity}
\label{Section - Brezis-Ekeland-Nayroles principle for standard plasticity and viscoplasticity}

 Based on seminal papers from 
Brezis and Ekeland \cite{Brezis Ekeland 1976} and Nayroles\cite{Nayroles 1976}, 
we build now a minimum principle for the plasticity that we shall call in short the \textbf{BEN principle}. We consider a solid body $\Omega$ with elasto-plastic (or elasto-viscoplastic) material. Its response to a given loading during the time interval form $0$ to $T$ is given by its stress field  $\bm{\sigma}$ and its displacement field $\bm{u} $. $\bm{S}$ being the elastic compliance matrix, we have as usual
\begin{equation}
    \dot{\bm{\varepsilon}} = \dot{\bm{\varepsilon}}^e + \dot{\bm{\varepsilon}}^p 
                                        = \bm{S} \, \dot{\bm{\sigma}} + \dot{\bm{\varepsilon}}^p 
\label{additive strain decomposition}
\end{equation}
We state that:
  \begin{itemize}
\item An evolution path $t \mapsto (\bm{\sigma} (t), \bm{u} (t))$  is said \textbf{admissible} if it satisfies
  the equilibrium equations, the kinematical conditions and the initial conditions
 \vspace{0.1cm}
 \item The \textbf{natural path} is the admissible path for which
 $\forall t \in \rbrack 0, T \lbrack$  we satisfy the normality law $\dot{\bm{\varepsilon}}^p \in \partial  \varphi (\bm{\sigma}) $
 with the potentials $\varphi (\bm{\sigma})  = I_{K} (\bm{\sigma}) $ and $\varphi^* (\dot{\bm{\varepsilon}}^p)$. Combining with (\ref{additive strain decomposition}), we have 
 $$\nabla \dot{\bm{u}} \in \bm{S} \dot{\bm{\sigma}} + \partial  \varphi (\bm{\sigma})
$$
or, because the subgradients are characterized by an inequation
$$\forall \bm{\sigma}' \in K, \qquad (\nabla \dot{\bm{u}} - \bm{S} \dot{\bm{\sigma}}) : (\bm{\sigma}'  - \bm{\sigma}) \leq 0
$$
we recover Hill's inequality.

\end{itemize}

\vspace{0.5cm}

\textbf{Non incremental BEN principle for standard plasticity and viscoplasticity.} \textit{The natural evolution path minimizes:}
$$\Pi(\bm{\sigma},\bm{u})  =  \int_{0}^{T}  \, \int_\Omega \,  \lbrack \varphi (\bm{\sigma}) + \varphi^* (\nabla \dot{\bm{u}} - \bm{S} \dot{\bm{\sigma}})  - \langle \bm{\sigma}, \nabla \dot{\bm{u}} - \bm{S} \dot{\bm{\sigma}}\rangle \rbrack \mbox{ d}\Omega \,\mbox{ dt}
$$
\textit{among all the admissible evolution paths and the minimum is zero.}

 \vspace{0.2cm}

\textbf{Proof.} Because of Fenchel's inequality
$$ \forall \bm{\sigma}',\dot{\bm{\varepsilon}}'^p \quad
\varphi (\bm{\sigma}') + \varphi^* (\dot{\bm{\varepsilon}}'^p) - \langle \bm{\sigma}',\dot{\bm{\varepsilon}}'^p \rangle \geq 0   
$$
using (\ref{additive strain decomposition}), integrating on the body and the time interval, we have
$$ \forall \bm{\sigma}',\bm{u}' \quad
      \Pi(\bm{\sigma}',\bm{u}') \geq 0
$$
On the other hand, the normality law 
$$\nabla \dot{\bm{u}} \in \bm{S} \dot{\bm{\sigma}} + \partial  \varphi (\bm{\sigma})
$$
is equivalent to the extremality condition
$$ \varphi (\bm{\sigma}) + \varphi^* (\nabla \dot{\bm{u}} - \bm{S} \dot{\bm{\sigma}})  = \langle \bm{\sigma}, \nabla \dot{\bm{u}} - \bm{S} \dot{\bm{\sigma}}\rangle
$$
Integrating on the body and the time interval, we find 
$$ \Pi(\bm{\sigma},\bm{u})  = 0
$$
that achieves the proof $\blacksquare$.

\vspace{0.2cm}

\textbf{Physical interpretation}. We start from the condition of minimum of the functional
\begin{equation}
     \int_{0}^{T}  \, \int_\Omega \,  \lbrack \varphi (\bm{\sigma}) + \varphi^* (\nabla \dot{\bm{u}} - \bm{S} \dot{\bm{\sigma}})  - \langle \bm{\sigma}, \nabla \dot{\bm{u}} - \bm{S} \dot{\bm{\sigma}}\rangle \rbrack \mbox{ d}\Omega \,\mbox{ dt} = 0
\label{condition of minimum for the BEN principle in plasticity}
\end{equation}
We time integrate by part the term in $\langle \bm{\sigma}, \bm{S} \dot{\bm{\sigma}}\rangle $. We move the term in $\langle \bm{\sigma}, \nabla \dot{\bm{u}} \rangle$ into the right hand member and  we transform it thanks to the principle of virtual powers. Finally we find\\
 \vspace{0.5cm}
$  \int_{0}^{T}  \, \int_\Omega \,  \lbrack \varphi (\bm{\sigma}) + \varphi^* (\nabla \dot{\bm{u}} - \bm{S} \dot{\bm{\sigma}})  \rbrack \mbox{ d}\Omega \,\mbox{ dt} \qquad\qquad\quad $ dissipated energy\\
 \vspace{0.2cm}
  $+  \int_\Omega \,  \lbrack 
  \frac{1}{2} \, \bm{\sigma} (T) : \bm{S} \bm{\sigma} (T)  - \frac{1}{2} \, \bm{\sigma} (0) : \bm{S} \bm{\sigma} (0) 
  \rbrack \mbox{ d}\Omega \qquad $ stored elastic energy\\
  \vspace{0.2cm}
  $ =  \int_{0}^{T}  \,  \lbrack \int_\Omega \,  \bar{\bm{f}} \cdot \dot{\bm{u}} \,  \mbox{d}\Omega 
  + \int_{\partial \Omega_1}  \bar{\bm{t}} \cdot \dot{\bm{u}} \, \mbox{d}S
  \rbrack  \,\mbox{ dt}\qquad\qquad\quad$ work of external forces

\vspace{0.2cm}

The physical meaning of this energy balance is that, for the natural path, the sum of the total dissipation energy and the total elastic energy stored during the loading (or lost if negative) is equal to the total work of the external forces ($\bar{\bm{f}} $ in the volume and $\bar{\bm{t}} $ on the part $\partial \Omega_1$ of the boundary of $\Omega$ where these forces are applied). 

\vspace{0.2cm}

\textbf{Remark 1.} We are learnt that the current state in plasticity depends on the load history. \\
In this sense, the BEN principle is counter-intuitive because all the states are calculated together and not step-by-step.

\begin{figure}[ht!]
\centering
\includegraphics[scale=.65]{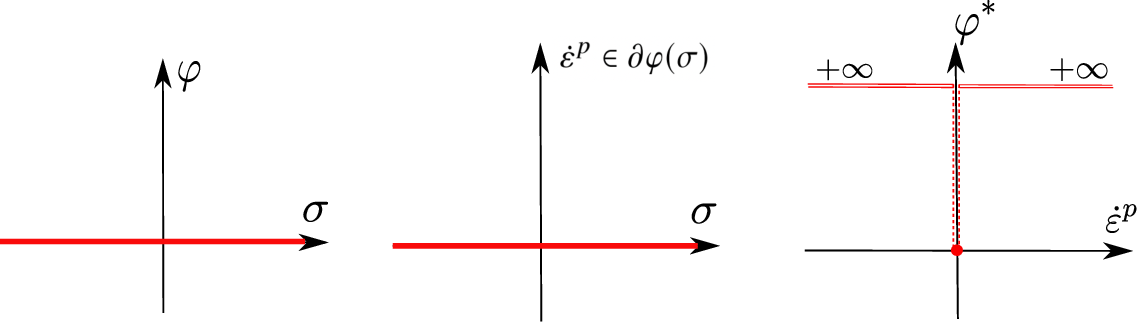}
\caption{Potentials for the 1D elasticity}
\label{fig Potentials for the 1D elasticity}
\end{figure}

\vspace{0.2cm}

\textbf{Remark 2.} Finally, we would like to discuss particular cases:
 
 \begin{itemize}
\item As we shall see later on, this kind of variational principle is devoted to the dissipative systems but it is worth to have a glance to the limit case of the  \textbf{elasticity} for which the plastic domain is the full stress tensor space $K = \mathbb{R}^6$ and the dual potentials are (Figure \ref{fig Potentials for the 1D elasticity})
$$ \varphi (\bm{\sigma}) = 0, \qquad 
\varphi^* (\dot{\bm{\varepsilon}}^p) =I_{\lbrace\bm{0} \rbrace} (\dot{\bm{\varepsilon}}^p)
$$
In the condition of minimum, the first term of the integrand of (\ref{condition of minimum for the BEN principle in plasticity}) vanishes. The value of the functional being null then finite, the second term must vanish and its argument as well, then we recover Hooke's law 
$$ \nabla \dot{\bm{u}} = \bm{S} \dot{\bm{\sigma}}
$$
The principle is degenerated but it works, even if there are simpler methods to  solve elasticity problems which are linear.
\vspace{0.2cm}
\item For the numerical applications, a drawback of this time-space formulation is that the evalation of the functional must be time consuming. If a step-by-step approach is prefered, we have in particular case an \textbf{incremental principle} by choosing  the loading time $\rbrack 0, T \lbrack$ reduced to a timestep. Hence the solving of the global problem on the whole loading interval is replaced by a sequence of incremental problems. Also intermediate strategies can be imagined with the integration over a few timesteps. 
\end{itemize}

\begin{figure}[ht!]
\centering
\includegraphics[scale=.75]{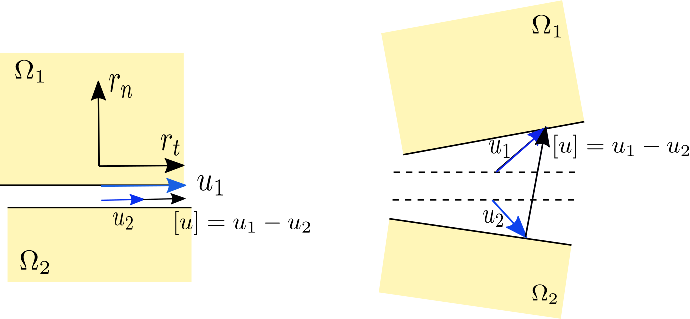}
\caption{Contact variables}
\label{FigureContactVariables}
\end{figure}

\vspace{0.2cm}

\section{BEN principle for Signorini unilateral contact}  
\label{Section BEN principle for Signorini unilateral contact}

\begin{figure}[ht!]
\centering
\includegraphics[scale=.65]{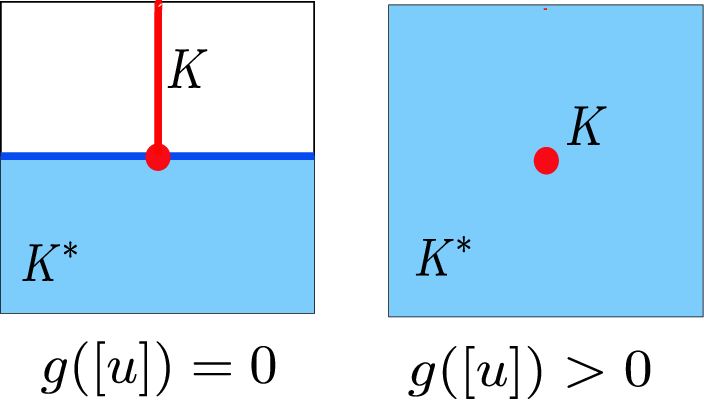}
\caption{Contact polar cones}
\label{FigureContactPolarCones}
\end{figure}

Let $\Omega_1$ and $\Omega_2$ be two bodies (see Figure \ref{FigureContactVariables}). We consider them initially in contact (on the left) or separated but closed then candidate to the contact (on the right). The displacement discontinuity between the body is
$$ \lbrack \bm{u} \rbrack = \bm{u}_1 - \bm{u}_2
$$
The contact reaction $\bm{r}$ subjected to $\Omega_1$ from $\Omega_2$ is the stress vector $\bm{t} = \bm{\sigma} \cdot \bm{n}$ if the reaction is continously distributed on the surface or a force if the contact is pointwise. The contact is checked with a gap function $g (\lbrack \bm{u} \rbrack)$ which may be the distance between the two points. The law of Signorini unilateral contact without friction is
$$ \textbf{if} \quad \mbox{the gap} \quad g (\lbrack \bm{u} \rbrack ) > 0, \qquad \bm{r}  = \bm{0} 
$$
$$ \textbf{else if} \quad g (\lbrack \bm{u} \rbrack ) = 0, \quad \bm{r} = r_n \bm{n}, \quad r_n \geq 0
$$
Nevertheless, we prefer to use a stronger formulation due to Jean-Jacques Moreau \cite{Moreau 1999} --called by him the viability lemma-- in which the law is expressed in terms of velocities (Figure  \ref{FigureContactPolarCones})
\begin{center}
\begin{tabular}{l l l}
   &  cone &  dual cone\\
 \textbf{if} $g (\lbrack \bm{u} \rbrack ) = 0$ & $ K = \lbrace \bm{r} = r_n \bm{n} , \; r_n \geq 0 \rbrace, $ & $
               K^* = \lbrace  - \lbrack \dot{\bm{u}} \rbrack \; \mbox{such that} \; - \lbrack \dot{u}_n \rbrack \leq 0 \rbrace$\\
\textbf{else if}  $g (\lbrack \bm{u} \rbrack ) > 0, $ & $K = \lbrace \bm{0} \rbrace, $ & $ K^* = \mathbb{R}^3$\\
 \end{tabular}
\end{center}

In terms of normality law, the potentials are the indicator functions of the dual cones
\begin{equation}
   \varphi (\bm{r}) = I_K  (\bm{r}) , \qquad 
   \varphi^* (- \lbrack \dot{\bm{u}} \rbrack  ) = I_{K^*}  (- \lbrack \dot{\bm{u}} \rbrack  ) 
\label{phi (r) = I_K (r) & phi^* (- (dot(u))) = I_K^* (- (dot(u))) }
\end{equation}
Mimicking what we did for the plasticity and viscoplasticity, we are able to state the corresponding variational formulation:

\vspace{0.2cm}

\textbf{Non incremental BEN principle for or elasticity with unilateral contact.} \textit{The natural evolution path minimizes:}
$$\Pi(\bm{\sigma},\bm{u})  =  \int_{0}^{T}  \, \int_{\partial \Omega_2} \,  \lbrace \varphi (\bm{r}) + \varphi^* (- \lbrack \dot{\bm{u}} \rbrack )  
+\lbrack \dot{\bm{u}} \rbrack \cdot \bm{r}  \rbrace \mbox{ d}S \,\mbox{ dt}
$$
\textit{among all the admissible evolution paths and the minimum is zero.}

 \vspace{0.2cm}
 
 The minimum which is zero is reached for a finite value of the integrand then we obtain an equivalent formulation by cancelling the indicator functions  in the functional and introducing the cone conditions as constraints of the minimization problem. 
 
 \vspace{0.2cm}

\textbf{Non incremental BEN principle for or elasticity with unilateral contact, version 2.} \textit{The natural evolution path minimizes:}
$$ \Pi(\bm{\sigma},\bm{u})  =  \int_{0}^{T}  \, \int_{\partial \Omega_2} \,  \lbrack \dot{\bm{u}} \rbrack \cdot \bm{r}  \mbox{ d}S \,\mbox{ dt}
$$
\textit{among all the admissible evolution paths such that 
$$ \bm{r} \in K , \qquad - \lbrack \dot{\bm{u}} \rbrack \in K^*
$$
and the minimum is zero.}

 \vspace{0.2cm}
 
The minimum condition leads to
$$\forall t \in \rbrack 0, T \lbrack, \qquad K \ni \bm{r} \perp - \lbrack \dot{\bm{u}} \rbrack  \in K^* 
$$ 
that is a \textbf{cone complementary problem}

As the elastic behaviour is linear, it is convenient to use Hooke's law and the admissibility conditions in order to reduce the size the problem. Indeed,  there exists an affine functional 
$$ \bm{r} \mapsto -  \lbrack \dot{\bm{u}} \rbrack  = \bm{A} \, \bm{r} + \bm{b} 
$$
solution of the problem of elasticity. In numerical simulation, there are standard techniques customarily used in the literature to obtain approximations of such a functional. Hence we can keep the contact reaction as unknowns and eliminate the other unknowns. Finally, the unilateral contact problem is reduced to
$$\forall t \in \rbrack 0, T \lbrack, \qquad   K \ni \bm{r} \perp  \bm{A} \, \bm{r} + \bm{b}  \in K^*
$$
called a \textbf{Linear Complementary Problem (LCP)} and for which efficient solvers of mathematical programing are available. As discussed at the end of the previous section, considerations as to the computer time in numerical simulations may lead to replace the global problem on the whole loading interval by a sequence of incremental problems of the same kind but with the interval reduced to a timestep. 

\section{Unified frameworks for dynamic dissipative systems}
\label{Section - Unified frameworks for dynamic dissipative systems}

There is a well-known geometry, that of Euclidean and Riemannian spaces, based on a symmetric 2-covariant tensor —the metric— to measure lengths. But there is another geometry, much less well known, that of symplectic spaces, based on an antisymmetric 2-covariant tensor —the symplectic form— to measure areas
$$ \omega (\bm{z}, \bm{z}') = \bm{z}^T \bm{ J}\, \bm{z}', \quad  \bm{J} = - \bm{J}^T 
$$
The origin of this geometry is to model the dynamics of reversible systems. We are working in $X \times Y$, the phase space of the Physicist, of which the elements are represented in local charts by the variables
$$\bm{z}  =\left[ \begin{array} {c}
                       \bm{\xi}  \\
                       \bm{\eta} \\
                    \end {array} \right]
$$
where $\bm{\xi}$ are the degrees of freedom —which may be also fields— and $\bm{\eta}$ are the associated dynamical momenta. It is equipped with the canonical symplectic form
\begin{equation}
\omega (\bm{z}, \bm{z}') = \bm{z}^T \bm{J} \,  \bm{z}'
                 =  \left[ \bm{\xi} , \bm{\eta}  \right]
                    \left[ \begin{array} {cc}
                       \bm{0}  & \bm{I}   \\
                     - \bm{I}  & \bm{0}  \\
                    \end {array} \right]\,
                    \left[ \begin{array} {c}
                       \bm{\xi} '  \\
                       \bm{\eta} ' \\
                    \end {array} \right] = \langle \bm{\xi} , \bm{\eta}' \rangle - \langle \bm{\eta}, \bm{\xi}' \rangle
\label{canonical symplectic form}
\end{equation}
defined by this antisymmetric matrix $\bm{J}$. It allows to construct the symplectic gradient, obtained by composition of $\bm{J}$ and the classical gradient
$$\dot{\bm{z}} = \nabla^\omega H (\bm{z} , t) = \bm{J} \,\nabla_{\bm{z} } H (t, \bm{z} )
$$
also often called Hamiltonian vector field in the literature and denoted $X_H$. In a mechanical point of view, it restitues the equation of motion in the form of the canonical equations
$$ \dot{\bm{\xi} }   =   \nabla_{\bm{\eta} } H, \qquad 
 \dot{\bm{\eta} } = - \nabla_{\bm{\xi} } H
$$
that are equivalent to the Euler-Lagrange equations of Hamilton least action principle claiming that for the natural evolution of a reversible system the action 
$$ A (\bm{\xi} )= \int^{T}_0 L (\bm{\xi}, \dot{\bm{\xi}}, t) \,\mbox{ d} t
$$
is stationary among all the paths $t \mapsto \bm{\xi} (t)$ with fixed extremities, but take care that it fails for dissipative systems which are our target. 

Nowadays, there are several unified frameworks for dissipative dynamical systems.
\begin{itemize}
\item Firstly, Morrison's metriplectic systems (see \cite{Barbaresco 2022}, \cite{Coquinot 2020}, \cite{Materassi 2017}, \cite{Morrison 1986}), developed further by Grmla and Öttinger (see \cite{Grmela 1997}, \cite{Ottinger 1997}) as GENERIC systems, with the Hamiltonian term for the reversible effects and Onsager's entropy term
$$ \dot{\bm{z}} = \bm{J} \,\nabla_{\bm{z} } H (\bm{z} ) 
                        + \bm{K}   \,\nabla_{\bm{z} } S (\bm{z} ), \qquad 
                        \bm{K} = \bm{K}^T > 0
$$                        
A variational formulation of GENERIC can be found in \cite{Manh Hong Duong 2013}. 
\item Introduced by \cite{Brockett 1977} and  \cite{van der Schaft 1984}, a similar approach but simpler is the one of port-Hamiltonian system with only one function but an extra operator $\bm{R}$ to model the resistive effects
$$ \dot{\bm{z}} = (\bm{J}  - \bm{R})\,\nabla_{\bm{z} } H (\bm{z} )
$$

\item One can mention also the rate-independent systems proposed by Mielke and Theil (\cite{mielketh99},  \cite{mielke}, \cite{MR06b}). Without entering into details, besides the energy function $E$, there is a dissipation potential $\phi$, 1-homogeneous with respect to the velocities
$$ \mbox{stability condition:} \qquad \nabla_{\bm{\xi}} E (\bm{\xi}) \cdot \bm{w} + \phi (\bm{w}) \geq  0, \; \forall \bm{w}
$$
$$ \mbox{power balance:} \;\; \nabla_{\bm{\xi}} E (\bm{\xi}) \cdot \dot{\bm{\xi}} + \phi (\dot{\bm{\xi}} ) = 0
$$
\item Finally, our formulation of Hamiltonian inclusions, initially proposed by Buliga \cite{bham} and developed later with the author \cite{SBEN}. It is inspired from both Morrison and Mielke approaches.  In addition to the Hamiltonian term, we have the symplectic gradient of a dissipation potential —derivating with respect to the velocities — but not necessarily 1-homogeneous or even homogeneous
$$ \dot{\bm{z}} = \bm{J} \,\nabla_{\bm{z} } H (\bm{z} ) +  \bm{J} \,\nabla_{\dot{\bm{z}} } \phi (\dot{\bm{z}} )
$$
\end{itemize}

Now, we describe in detail the last formulation in the more general framework of nonsmooth mechanics.  For the evolution of a dissipative dynamical system, we suppose an additive decomposition of the velocity into reversible and irreversible parts: 
$$ \dot{\bm{z}} =  \dot{\bm{z}}_{R} + \dot{\bm{z}}_{I},\quad 
\dot{\bm{z}}_{R} =  \nabla^\omega H,  \quad 
\dot{\bm{z}}_{I}  =  \dot{\bm{z}}  -  \nabla^\omega H 
$$
where the reversible part of the velocity is given by the symplectic gradient of the Hamiltonian $H$
$$ \omega (\nabla^\omega H , \bm{\zeta}) = \lim_{\epsilon \to 0} \frac{1}{\epsilon} 
      (H (t, \bm{z} + \epsilon \, \bm{\zeta}) 
      - H (t, \bm{z}) )
$$
Let $\phi: X \times Y \rightarrow \mathbb{R}\cup \left\{+\infty\right\}$ be a convex lower semicontinuous  dissipation potential.  The \textbf{symplectic subdifferential} of $\phi$ at $\bm{z}$ where $\phi$ has a finite value is the set
\begin{eqnarray}
    \partial^{\omega} \phi (\dot{\bm{z}})  =  
\{ 
\dot{\bm{z}}_I \in X \times Y \mbox{ : } \forall  \dot{\bm{z}}'' \in X \times Y \quad \nonumber \\ 
\phi(\dot{\bm{z}}+\dot{\bm{z}}'') - \phi(\dot{\bm{z}}) \geq
\omega (\dot{\bm{z}}_I , \dot{\bm{z}}'') \} 
\label{dssub}
\end{eqnarray}
The elements of the symplectic subdifferential are called symplectic subgradients and satisfy the so-called Hamiltonian inclusion
\begin{equation}
   \dot{\bm{z}}_I \in \partial^{\omega} \phi (\dot{\bm{z}}) 
   \label{z' in partial^omega F (z)}
\end{equation}In a physical point of view, it is the dissipative constitutive law. 
The \textbf{symplectic polar }of $\phi$ is the function 
\begin{equation}
\phi^{*\omega}(\dot{\bm{z}}_I) =  \sup \left\{ \omega(\dot{\bm{z}}_I,\dot{\bm{z}}) - \phi(\dot{\bm{z}}) \mbox{ : } \dot{\bm{z}} \in X \times Y \right\}
\label{dspolar}
\end{equation}
By construction of the symplectic polar, the \textbf{symplectic Fenchel inequality} holds
\begin{equation}
\forall  \dot{\bm{z}}, \dot{\bm{z}}_I \in X \times Y, \qquad
  \phi(\dot{\bm{z}}) + \phi^{*\omega}(\dot{\bm{z}}_I) - \omega(\dot{\bm{z}}_I, \dot{\bm{z}}) \geq 0
\label{symplectic Fenchel inequality}
\end{equation}
The equality in the previous relation is reached when $\dot{\bm{z}}_I$ is a symplectic subgradient of $\phi$ at $\bm{z}$. Indeed, it has been proved  in \cite{SBEN} that the dissipative constitutive law is equivalent to the \textbf{extremality condition}
\begin{equation}
   \dot{\bm{z}}_I \in \partial^{\omega} \phi (\dot{\bm{z}})
  \quad \Leftrightarrow \quad 
 \phi(\dot{\bm{z}}) + \phi^{*\omega}(\dot{\bm{z}}_I) - \omega(\dot{\bm{z}}_I,\dot{\bm{z}}) = 0
\label{extremality <-> dot(z)_I in partial^omega F(dot(z))}
\end{equation} 
In such case, we say that the couple $(\dot{\bm{z}}_I,\dot{\bm{z}}) $ is extremal.  

\section{A symplectic minimum variational principle for dynamical dissipative media}
\label{Section - A symplectic minimum variational principle for dynamical dissipative media}

Inspiring from the original idea of Brezis and Ekeland\cite{Brezis Ekeland 1976} and Nayroles\cite{Nayroles 1976}, the author proposed with Buliga in \cite{SBEN} a symplectic version in the framework of dissipative dynamical systems. The loading being specified during the time interval from $0$ to $T$, an evolution path $\left[0, T\right] \rightarrow X \times Y: t \mapsto \bm{z} (t)$  in the $X \times Y$ is called an \textbf{admissible path} if it satisfies the initial and boundary conditions. The \textbf{natural path} is the admissible path for which the dissipative constitutive law (\ref{z' in partial^omega F (z)}) is satisfied at every time during the load interval. 

\vspace{0.5cm}

\textbf{Symplectic Brezis-Ekeland-Nayroles principle (SBEN).} \textit{The natural evolution path minimizes:}
\begin{equation}
    \Pi(\bm{z})  =  \int_{0}^{T} \left[\phi(\dot{\bm{z}}) 
                                              + \phi^{*\omega}(\dot{\bm{z}}  - \nabla^\omega H) -  
                                              \omega (\dot{\bm{z}}  - \nabla^\omega H, \dot{\bm{z}} ) 
                                       \right] \mbox{ dt} 
\label{Functional of the SBEN principle}
\end{equation}
\textit{among all the admissible evolution paths and the minimum is zero.}

 \vspace{0.2cm}

This functional is built from the dissipation potential, the Hamiltonian for the reversible behavior and the symplectic form for the dynamics. The proof is similar to that of the non incremental BEN principle for standard plasticity and viscoplasticity of Section \ref{Section - Brezis-Ekeland-Nayroles principle for standard plasticity and viscoplasticity}, using the symplectic Fenchel inequality (\ref{symplectic Fenchel inequality}) and the extremality condition of (\ref{extremality <-> dot(z)_I in partial^omega F(dot(z))}). The functional is not convex but there is (at least partial) convexity, that is favourable for the convergence of the minimization algorithm. 

\section{Application : standard plasticity and viscoplasticity in dynamics}
\label{Section - Application : standard plasticity and viscoplasticity in dynamics}

The elements of the space $X$ are the couples $\bm{\xi} =  (\bm{u}, \bm{\varepsilon}^p) $ of fields of displacement and plastic strain. The corresponding dynamical momenta are denoted $\bm{\eta} = (\bm{p} , \bm{\pi}) $. The dual pairing is of the form
$$ \left\langle \dot{\bm{x}} , \dot{\bm{y}} \right\rangle 
 = \left\langle \dot{\bm{u}} , \dot{\bm{p}} \right\rangle + \left\langle \dot{\bm{\varepsilon}}^p , \dot{\bm{\pi}} \right\rangle 
= \int_\Omega \dot{\bm{u}} \cdot \dot{\bm{p}}  \, d\Omega
  + \int_\Omega \dot{\bm{\varepsilon}}^p : \dot{\bm{\pi}}  \, d\Omega
$$
The Hamiltonian of the body is 
$$     H (t, \bm{z}) 
    = \int_{\Omega} \left\lbrace \dfrac{1}{2 \rho} \parallel \bm{p} \parallel^ 2 
    + w\, (\nabla \bm{u} - \bm{\varepsilon}^p) - \bm{f} (t)\cdot\bm{u} \right\rbrace  \, d\Omega \nonumber \\
              - \int_{\partial\Omega_1}  \bar{\bm{f}} (t)\cdot \bm{u}  \, d\Gamma \nonumber
$$
The first term is the kinetic energy, $w$ is  the quadratic elastic strain energy expressed in term of the compliance matrix $\bm{S}$
$$ w (\bm{\varepsilon}^e) =   \frac{1}{2} \, \bm{\varepsilon}^e : \bm{S}^{-1} \bm{\varepsilon}^e
$$
$\bm{f}$ is the volume force and $\bar{\bm{f}}$ is the surface force on the part $\partial\Omega_1$ of the boundary, the displacement field being equal to an imposed value $\bar{\bm{u}}$ on the remaining part $\partial\Omega_0$. The stress field is given by the elasticity law
\begin{equation}
\label{elasticity law}
      \bm{\sigma} = \frac{\partial w}{\partial \bm{\varepsilon}} \,  (\nabla \bm{u} - \bm{\varepsilon}^p)
                          = \bm{S}^{-1} \, : (\nabla \bm{u} -\bm{\varepsilon}^p )
\end{equation}
We consider the canonical symplectic form (\ref{canonical symplectic form}). As the degrees of freedom are fields, the symplectic gradient of the Hamiltonian must be taken in the sense of the Calculus of Variation, then it is expressed in terms of \textbf{functional derivatives} denoted by $D$
$$ \nabla^\omega H = ((D_{\bm{p}} H,D_{\bm{\pi}} H), (- D_{\bm{u}} H, - D_{\bm{\varepsilon}^p} H))
= ( ( \bm{p} / \rho, \bm{0} ) , ( \nabla \cdot \bm{u} + \bm{f} , \bm{\sigma} ) )
$$
Thus one has
$$ \dot{\bm{z}}_I = \dot{\bm{z}} - \nabla^\omega H  = ( ( \bm{u}_I , (\bm{\varepsilon}^p)_I ) , ( \bm{p}_I, \bm{\pi}_I ) )
$$
with
$$ \dot{\bm{u}}_I =\dot{\bm{u}} -  \bm{p} / \rho , \quad
     (\dot{\bm{\varepsilon}}^p)_I = \dot{\bm{\varepsilon}}^p , \qquad 
   \dot{\bm{p}}_I = \dot{\bm{p}} - \nabla \cdot \bm{\sigma} - \bm{f} , \quad
     \dot{\bm{\pi}}_I = \dot{\bm{\pi}} - \bm{\sigma}
$$
The symplectic polar function of  $\phi$ reads
$$\phi^{*\omega} (\dot{\bm{z}}_I) 
        = \sup \left\{ \left\langle \dot{\bm{u}}_I, \dot{\bm{p}}\right\rangle
                     + \left\langle (\dot{\bm{\varepsilon}}^p)_I, \dot{\bm{\pi}}\right\rangle
                     - \left\langle \dot{\bm{u}}, \dot{\bm{p}}_I\right\rangle
                     - \left\langle \dot{\bm{\varepsilon}}^p, \dot{\bm{\pi}}_I\right\rangle
                     - \phi(\dot{\bm{z}}), \quad  \dot{\bm{z}} \in X \times Y \right\} 
$$
Now it is a crucial turning point. To recover the standard plasticity, we make the strong hypothesis that $\phi$ depends explicitly only on $\dot{\bm{\pi}}$ through $\varphi$
\begin{equation}
      \phi(\dot{\bm{z}}) = \varphi (\dot{\bm{\pi}})
\label{phi(dot(z)) = varphi(dot(pi))}
\end{equation}
In other words, $\dot{\bm{u}}, \dot{\bm{\varepsilon}}^p, \dot{\bm{p}}$ are ignorable in $\phi$. Therefore $\phi^{*\omega} (\dot{\bm{z}}_{I}) = \varphi^* (\dot{\bm{\varepsilon}}^p)$ (polar function of $\varphi$) under these 3 conditions:
\begin{itemize}
\item $\bm{p} = \rho \dot{\bm{u}}$ is the linear momentum, 
\item the balance of linear momentum is satisfied
\begin{equation*}
   \nabla \cdot \bm{\sigma} + \bm{f} = \dot{\bm{p}} = \rho \ddot{\bm{u}}\quad \mbox{on} \quad\Omega
\end{equation*}
\item and $\dot{\bm{\pi}} = \bm{\sigma}$, that reveals  the physical meaning of $\bm{\pi}$. 
\end{itemize}
Otherwise, its value is infinite. How do we use this fact ? As the minimum zero cannot be reached for infinite values, the three previous conditions are introduced as constraints in the minimization. Eliminating the momenta thanks to the two last relations, we obtain the variational principle:

\vspace{0.5cm}

\textbf{SBEN principle for standard plasticity and viscoplasticity in dynamics.} \textit{The natural evolution path minimizes:}
$$\Pi(\bm{\sigma},\bm{u})  =  \int_{0}^{T}  \, \int_\Omega \,  \lbrack \varphi (\bm{\sigma}) + \varphi^* (\nabla \dot{\bm{u}} - \bm{S} \dot{\bm{\sigma}})  - \langle \bm{\sigma}, \nabla \dot{\bm{u}} - \bm{S} \dot{\bm{\sigma}}\rangle \rbrack \mbox{ d}\Omega \,\mbox{ dt}
$$
\textit{among all the admissible evolution paths satisfying the balance of linear momentum}
$$ \nabla \cdot \bm{\sigma} + \bm{f}  = \rho \ddot{\bm{u}}
$$
\textit{and the minimum is zero.}
 
 \vspace{0.2cm}

The quasi-static limit case of Section \ref{Section - Brezis-Ekeland-Nayroles principle for standard plasticity and viscoplasticity} is obtained by neglecting the inertia forces. Invoking the strong hypothesis (\ref{phi(dot(z)) = varphi(dot(pi))})  to recover the classical plasticity means that we insert the "statical dissipative constitutive law"
$$\dot{\bm{\varepsilon}}^p \in \partial  \varphi (\bm{\sigma}) 
$$
 in the dynamical framework. Giving up this hypothesis paves the way to a much more general modeling  by a  "full dynamical constitutive law"
$$ \dot{\bm{z}}_{I} \in \partial^\omega \phi (\dot{\bm{z}}) \qquad \Leftrightarrow \qquad
   ((\dot{\bm{u}}_{I},\dot{\bm{\varepsilon}}^p),(\dot{\bm{p}}_{I}, \dot{\bm{\pi}}_{I})) 
   \in \partial^\omega 
   \phi ((\dot{\bm{u}},\dot{\bm{\varepsilon}}^p),(\dot{\bm{p}}, \dot{\bm{\pi}}))
$$

Numerical applications of this principle have showed the relevancy of the variational approach. In particular,  \cite{Cao 2020} and \cite{Cao 2021a} are devoted to the numerical simulation of the statics of thick tubes and of plates and \cite{Cao 2021b} to the dynamics of thick and thin tubes. For the dynamic plasticity, there is a lot of experimental and numerical papers but a lack of analytical solution. Then we have been led to develop a closed-form solution for our own use in \cite{Cao 2022}.

\section{Other applications}
\label{Section - Other applications}

The previous approach has been extended in \cite{Cao 2022b} to the \textbf{plasticity in finite strains}. Because the target applications in the future concern solids, we opt for a \textbf{Lagrangian specification} of the matter flow in the sense that the field are expressed in terms of the time and the initial position $\bm{x}_0$ of the material particle of which we follow the trajectory (as opposed to the Eulerian specification for which the fields are expressed in terms of the time and the current position $\bm{x}$ at this moment). In the literature we can distinguish the approaches relied on the additive and multiplicative decompositions. 

The first one is based on the \textbf{additive decomposition} of the Green-Lagrange measure $\textcolor{black}{\bm{E} = \bm{E}^e + \bm{E}^p}$ proposed by Green and Nagdhi \cite{Green Naghdi 1965}. This formalism is convenient to study the geometric effects of the second order for the beam, plate and shells in situations where the deformations remains small while the rotations and displacements are large \cite{Nguyen 2012}. The elements of the space $X$ are couples 
$$\bm{\xi} = (\bm{x},\bm{E}^p) 
$$

For materials undergoing large deformations, the additive decomposition seems not to be adapted and it can be replaced by the \textbf{multiplicative decomposition} of the deformation gradient $\bm{F} = \bm{F}^e \bm{F}^p$ introduced by Lee (\cite{Lee 1967}, \cite{Lee 1969}), where the elastic part of the deformation $\bm{F}^e$ is obtained by unloading all infinitesimal neighborhoods of points of the body. The elements of the space $X$ are couples 
$$\bm{\xi} = (\bm{x},\bm{F}^p ) 
$$ 
For each of these two approaches, we proposed a specific variational formulation. 

We also extended the SBEN principle to the models of Navier-Stokes equations and Bingham fluids in \cite{SBEN_NS}. For flows, we consider the \textbf{Eulerian specification} where the degree of freedom is the field of current position $\bm{x}$ and the dynamical momentum is the Eulerian linear momentum $\bm{p} = \rho \, \bm{v}$. Inspiring from Hilbert-Einstein action in General Relativity \cite{GR, Soper 1976}, we represent the fluid motion by $\bm{x}_0 = \kappa (t, \bm{x}) = \kappa (\bm{X})$. To calculate the functional derivatives, we use a special form of the calculus of variation by replacing the original field $\bm{x}_0$ by its first jet prolongation $j^1 \bm{x}_0$, a map from the body $\Omega$ into the jet bundle $J^1 (\Omega, \mathbb{R} ^3)$ such that $(j^1 \bm{x}_0) (\bm{X}) = (\bm{X}, \bm{x}_0 (\bm{X}),   (\nabla_{\bm{X}} \bm{x}_0) (\bm{X}))$, that leads to perform variations not only on the field $\bm{x}_0 $ and its derivatives but also on the variable $\bm{X}$. 

The equations of our approach are covariant with respect to the Galilei principle of relativity \cite{AffineMechBook}, taking into account the gravity $\bm{g}$ and the Coriolis vector $\bm{\Omega}$ useful for the applications to environmental fluids (atmosphere, oceans, Earth's interior). As usual, $p$ is the pressure and we use the material derivative $D / Dt = \partial / \partial t + v \cdot \nabla $. The convex dissipation potential $\varphi$ is supposed to depend explicitly only on the Eulerian velocity $\bm{v}$ through the strain rate. For the Navier-Stokes model, it is quadratic while for the Bingham fluid it is nonsmooth. 

\vspace{0.5cm}

\textbf{SBEN principle for compressible Navier-Stokes equations and Bingham fluids.}
\textit{The natural evolution path} $t \mapsto (\kappa_t, \bm{v})$ \textit{minimizes the functional}
\begin{eqnarray}
 \Pi [\kappa, \bm{v}]  & = &
    \int^T_0 \lbrace \varphi (\bm{v}) 
                              + \varphi^{*} \left( - \rho\,\frac{D \bm{v}}{D t}  - \nabla p  + \rho\,\left(\bm{g} - 2\,\bm{\Omega} \times \bm{v} \right) \right) \nonumber\\
         &   &     + \int_{\Omega_t}  
\left[ \rho\,\frac{D \bm{v}}{D t}  + \nabla p  - \rho\, \bm{g}  \right] \cdot \bm{v} \,
 \mbox{d}^3 x  \rbrace\ \mbox{d}t 
\label{Pi (kappa, v)}
\end{eqnarray}
\textit{among all the admissible evolution paths, and the minimum is zero.}

\vspace{0.2cm}

A corresponding principle is obtained for the limit case of \textbf{incompressible flows} \cite{SBEN_NS}. For a flow, the term of the last line in the functional (\ref{Pi (kappa, v)}) is the sum of the velocity head, pressure head and elevation head losses due to dissipation during the interval from $0$ to $T$.  For the limit case of \textbf{inviscid flows}, the potential of dissipation $\varphi$ vanishes and its polar function $\varphi^*$ has a finite value equal to zero if its argument vanishes, \textit{i.e.} Euler's equations, then the SBEN principle claims that the total head loss is zero, that is the expression of \textbf{Bernoulli's principle}. 

\section{Bipotentials}
\label{Section - Bipotentials}

There is a class of constitutive laws which escape from attempts of getting them into the mold of the approach by convex potentials, the \textbf{non associated  constitutive laws}. For them, we proposed in \cite{saxfeng, sax CRAS 92}  the bipotentials that allow extending the  classical variational principles of minimum in a simple way. A rigorous mathematical formalism was proposed in (\cite{Buliga 2008}, \cite{Buliga 2010}). A bipotential is a function $b: X \times Y \rightarrow \mathbb{R} \cup \left\lbrace +\infty \right\rbrace$ such that:
\begin{itemize}
    \item [(i)] $b$ is bi-convex (\textit{i.e.} convex with respect to each of its arguments) and bi-lower semicontinuous
    \item  [(ii)]   $\forall \bm{x}' \in X, \; \bm{y}' \in Y,\qquad b(\bm{x}', \bm{y}') \geq \left\langle \bm{x}', \bm{y}' \right\rangle$
    \item   [(iii)]   If one of the following relations is true, the other ones also are
    \begin{eqnarray}
        b(\bm{x}, \bm{y}) = \left\langle \bm{x}, \bm{y} \right\rangle 
                \qquad \Leftrightarrow \qquad  \nonumber  
                \bm{y} \in \partial b (\bullet, \bm{y}) (\bm{x})
                \qquad \Leftrightarrow \qquad  
                \bm{x} \in \partial b (\bm{x}, \bullet) (\bm{y})  \nonumber
    \end{eqnarray}
   The meaning of the second relation is that we differentiate with respect to $\bm{x}$ at constant $\bm{y}$, then $\bm{y}$ is an implicit   function of $\bm{x}$. 
\end{itemize}
In the particular event of associated laws, the bipotential is separated
$$ b(\bm{x}, \bm{y}) = \varphi(\bm{x}) + \varphi^* ( \bm{y})
$$
where the dissipation potential $\varphi$ and its Fenchel polar $\varphi^*$ are convex and lower semicontinuous. For such separated bipotentials, the equivalence of the three relations of (iii) is true as a consequence of (i) and (ii), as discussed in Section \ref{Section - A smidgen of convex analysis}. It is not so for a non separated bipotential, reason for which it is an axiom of the definition of bipotentials that must be proved on a case-by-case basis.

\begin{figure}[ht!]
	\centering
	\includegraphics[scale=0.50]{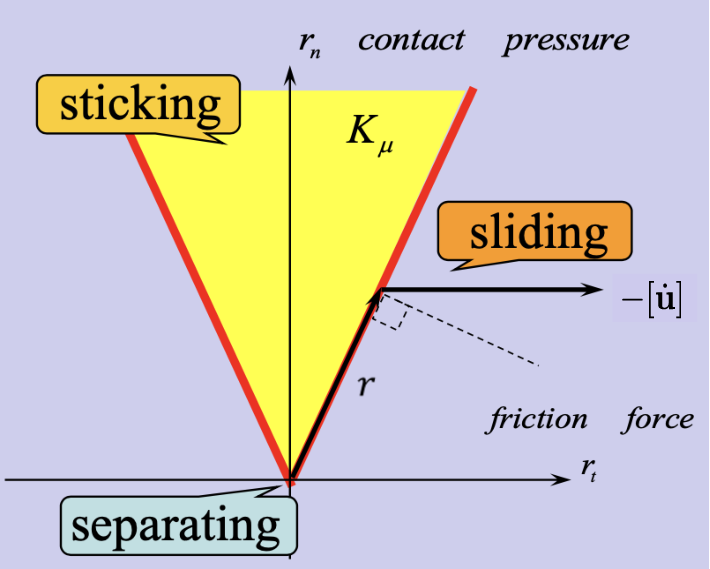}
	\caption{Unilateral contact with Coulomb friction law}
	\label{Figure Coulomb law}
\end{figure}

Maybe the oldest non associated law is the one of Coulomb’s frictional contact (Figure \ref{Figure Coulomb law}). \textbf{Coulomb's cone} is defined by:
$$ K_\mu = \left\lbrace \bm{r} \in \mathbb{R}^3 \quad \mbox{such that} \quad  
                      f (\bm{r}) = \parallel \bm{r}_t \parallel  - \mu\, r_n \leq 0
           \right\rbrace\ ,
$$
where $\mu > 0$ is the friction coefficient. The law of unilateral contact with Coulomb's dry friction is:

% (Figure Coulomb law)

\vspace{0.5cm}

\begin{tabular}{|lll|}
\cline{1-3}
 & & \\
 \textbf{if} &  $\bm{r} \in \mbox{int} (K_\mu)$ \textbf{then} & ! sticking\\
                       &  $\lbrack \dot{\bm{u}} \rbrack = \bm{0}$ &  \\
 \textbf{elseif}   &  $\bm{r}\in \partial K_\mu - \left\lbrace \bm{0} \right\rbrace $ \textbf{then} & ! sliding \\
                       &  $ \lbrack \dot{u}_n \rbrack = 0\quad \mbox{and}\quad 
                               \exists \lambda \geq 0 \quad \mbox{such that} \quad
                               \lbrack \dot{\bm{u}}_t\rbrack = - \lambda \, \dfrac{\bm{r}_t}{\parallel \bm{r}_t \parallel} $ & \\
 \textbf{if} &  $\bm{r} = \bm{0}$ & ! apex \\
                 &  $ \lbrack \dot{u}_n \rbrack \geq 0$ & ! no contact \\
 & & \\
 \cline{1-3}
\end{tabular}

\vspace{0.5cm}
Moreover, there is no velocity associated to reactions outside of Coulomb's cone $K_\mu$. Its boundary $\partial K_\mu$ is called the \textbf{sliding surface}. The law is non associated because the relative velocity vector is not normal to the sliding surface.  As the graph is not monotone, it cannot be generated by a convex potential. Nevertheless, it can be generated by a bipotential that we construct by an heuristic method. The idea is to majorize the dual pairing on the graph of the constitutive law. We have to do it just what is necessary. In other words, the majorant is the supremum and must be reached for couples of the graph, in fact those which are extremal. If the couple $( - \lbrack \dot{\bm{u}} \rbrack,  \bm{r})$ belongs to the graph
\begin{equation}
  \lbrack \dot{u}_n \rbrack \geq 0  \ ,
\label{dot(u)_n geq 0} 
\end{equation}
and because  $\bm{r} \in K_\mu $, $r_n \geq 0$ then:
$$ - \lbrack \dot{\bm{u}} \rbrack \cdot \bm{r} =
     - \lbrack \dot{u}_n \rbrack \; r_n - \lbrack \dot{\bm{u}}_t \rbrack\cdot \bm{r}_t
     \leq - \lbrack \dot{\bm{u}}_t \rbrack\cdot \bm{r}_t
     \leq \parallel - \lbrack \dot{\bm{u}}_t \rbrack \parallel\,\parallel\bm{r}_t \parallel\ ,
$$
and once again because:
\begin{equation}
   \bm{r} \in K_\mu\ ,
\label{r in K_mu} 
\end{equation}
we obtain the majorization:
$$ - \lbrack \dot{\bm{u}} \rbrack \cdot \bm{r} \leq  \mu\,r_n \parallel - \lbrack \dot{\bm{u}} \rbrack \parallel\ .
$$
We can verify that the majorant is reached for the extremal couples. It provides the finite part of the bipotential that we complete by indicator functions associated to the admissibility conditions, then the expression of the bipotential
\begin{equation}
\label{bipo Coulomb} 
   b (-\lbrack \dot{\bm{u}}\rbrack, \mathbf{r}) \ =  \mu r_{n} \parallel - \lbrack \dot{\bm{u}}_t \rbrack \parallel 
                                                          +   I_{K_{\mu}} (\mathbf{r}) 
                                                          +  I_{ K^{*}} (-\lbrack \dot{\bm{u}} \rbrack) 
\end{equation} 
where the cone $K^*$ was defined in Section \ref{Section BEN principle for Signorini unilateral contact}. The cornerstone inequality (ii) of the bipotential is verified by construction. It was done for that. Next, we have to verify the other properties, overall the equivalence (iii)  which is not automatic. In the limit case of non friction ($\mu = 0$), the friction cone $K_\mu$ degenerates into the cone $K$ of  Section \ref{Section BEN principle for Signorini unilateral contact} and the bipotential is separated, sum of the indicator functions (\ref{phi (r) = I_K (r) & phi^* (- (dot(u))) = I_K^* (- (dot(u))) }) of the dual cones of the Signorini unilateral contact. 

\section{Symplectic bipotentials and SBEN principle for non associated constitutive laws}
\label{Section - Symplectic bipotentials and SBEN principle for non associated constitutive laws}

As in Section \ref{Section - Unified frameworks for dynamic dissipative systems}, the space $X \times Y$ is equipped with the symplectic form $\omega$. For the non associated laws in the context of dynamics, we propose to define a symplectic bipotential as a function $\hat{b}: (X \times Y)^2 \mapsto \mathbb{R} \cup \left\lbrace +\infty \right\rbrace$ such that:
\begin{itemize}
    \item [(a)] $\hat{b} $ is bi-convex and bi-lower semicontinuous
    \item [(b)] $\forall \dot{\bm{z}}', \dot{\bm{z}}'_I \in Z,\qquad \hat{b} (\dot{\bm{z}}'_I, \dot{\bm{z}}' )                                                      \geq \omega (\dot{\bm{z}}'_I, \dot{\bm{z}}' )$
    \item [(c)] If one of the following relations is true, the other ones also are
    \begin{eqnarray}
     \hat{b} (\dot{\bm{z}}_I, \dot{\bm{z}}) = \omega (\dot{\bm{z}}_I, \dot{\bm{z}} )
                \quad \Leftrightarrow \quad \nonumber  
                \dot{\bm{z}}_I \in \partial^{\omega} \hat{b}  (\dot{\bm{z}}_I, \bullet) (\dot{\bm{z}})
                \quad \Leftrightarrow \quad  
                - \dot{\bm{z}} \in \partial^{\omega} \hat{b}  (\bullet, \dot{\bm{z}}) (\dot{\bm{z}}_I)   \nonumber  
    \end{eqnarray}
\end{itemize}
The minus sign in (b) results from the antisymmetry of $\omega$.

In \cite{de Saxce 2026}, we proposed a method to construct a symplectic bipotential  from a bipotential in the particular case of interest where $X=Y$ is a Hilbert space, then the dual pairing  $\langle \bullet, \bullet \rangle$ is a scalar product. Moreover the space $Z = X \times Y$ is dual with itself, according to
$$    \langle \langle (\bm{x},\bm{y}), (\bm{x}', \bm{y}') \rangle \rangle 
    \, = \, \langle \bm{x}, \bm{x}' \rangle + \langle \bm{y}, \bm{y}' \rangle 
$$
Introducing the continuous endomorphism $\bm{J}$  of Section \ref{Section - Unified frameworks for dynamic dissipative systems}, $ Z $ is a symplectic vector space for 
$$\omega(\bm{z}, \bm{z}') \, = \, \langle \langle \bm{z}, \bm{J} \, \bm{z}' \rangle \rangle 
                                            = \, \langle \bm{x}, \bm{y}' \rangle - \langle \bm{x}', \bm{y} \rangle 
 $$
Now we use an heuristic, focussing our attention on the particular case of associated laws for which the extremality condition (\ref{extremality <-> dot(z)_I in partial^omega F(dot(z))}) is
$$ \phi(\dot{\bm{z}}) + \phi^{*\omega} (\dot{\bm{z}}_I) - \omega( \dot{\bm{z}}_I, \dot{\bm{z}}) = 0
$$
in which now the last term is
$$ \omega (\dot{\bm{z}}_I, \dot{\bm{z}}) = 
          \langle \langle \dot{\bm{z}}_I, \bm{J} \, \dot{\bm{z}}\rangle \rangle 
$$
Comparing to the extremality condition of Section \ref{Section - A smidgen of convex analysis}, we show that the value of the symplectic polar is the classical polar with the following argument
$$\phi^{*\omega} (\dot{\bm{z}}_I) = \phi^* (- \bm{J} \,  \dot{\bm{z}}_I)
$$
On this basis, we come back to the general case of non associated laws and —by induction— we claim that 
$$ \hat{b} (\dot{\bm{z}}_I, \dot{\bm{z}}) = b (- \bm{J}\,\dot{\bm{z}}_I, \dot{\bm{z}}) 
$$
Of course, it is only an heuristic and we have to prove that if $b$ is a bipotential, $\hat{b}$ is a symplectic bipotential, 
that is demonstrated in \cite{de Saxce 2026}.

Now we come back to the calculus of variations.  The  key idea is to replace in the functional (\ref{Functional of the SBEN principle}) the sum of the potential of dissipation and its symplectic polar by a symplectic bipotential $\hat{B} $. 

\vspace{0.5cm}
 
\textbf{SBEN principle for symplectic bipotentials.} \textit{The natural evolution of the system minimizes the functional}
\begin{equation}
\Pi(\bm{z}) = \int_{0}^{T} \left\{ \hat{B}       
    (\dot{\bm{z}} - \nabla^\omega H , \dot{\bm{z}})
    - \omega (\dot{\bm{z}} - \nabla^\omega H, \dot{\bm{z}})  \right\} \mbox{ dt} 
\label{SBEN 2 Pi (z) =}
\end{equation}
\textit{among the admissible paths $t \mapsto \bm{z} (t) $ and the minimum is zero.} 

\begin{figure}[ht!]
	\centering
	\includegraphics[scale=1.00]{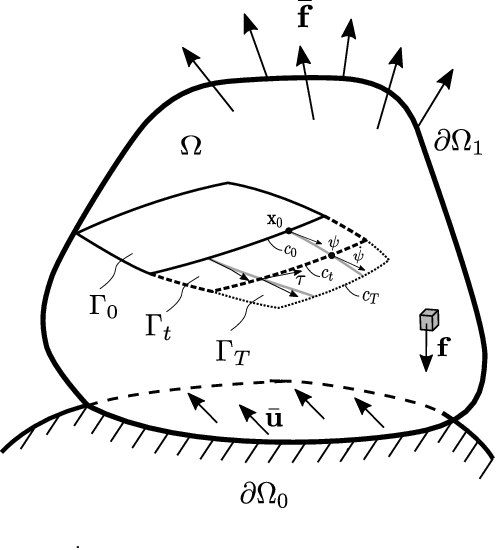}
	\caption{Crack Flow}
	\label{FigCrackFlow}
\end{figure}

\section{Application to the crack propagation with friction between the crack sides}
\label{Section - Application to the crack propagation with friction between the crack sides}

To illustrate, we consider the problem of extension of a crack in a brittle elastic material in small strain with possible friction contact between the crack sides (Figure \ref{FigCrackFlow}). The crack evolution is described by a time-parameterized family of surfaces $  t \mapsto \Gamma_t = \Gamma (t)$. The crack front $c_t$ at time $t$ is parameterized by the arc length $s$. The key-idea of the formulation proposed in \cite{de Saxce 2022} is to model the crack extension by a flow on the cracked surface $\Gamma_T \backslash \Gamma_0$ during the time interval 
\begin{equation}
\left[ 0, T \right] \times c_0 \rightarrow \Gamma_T \backslash \Gamma_0 : (t, \bm{x}_0) \mapsto \bm{x} = \bm{\psi} (t, \bm{x}_0)
\label{crack flow}
\end{equation}
The advantages are that it is easier working with fields living in a functional space than with surfaces and the approach is well suited for numerical applications. Then $\Gamma_T\backslash \Gamma_0 $ can be seen as the \textbf{flow of the vector field} $\dot{\bm{\psi}}$ (Eulerian, not Lagrangian). 

\begin{figure}[ht!]
	\centering
	\includegraphics[scale=0.30]{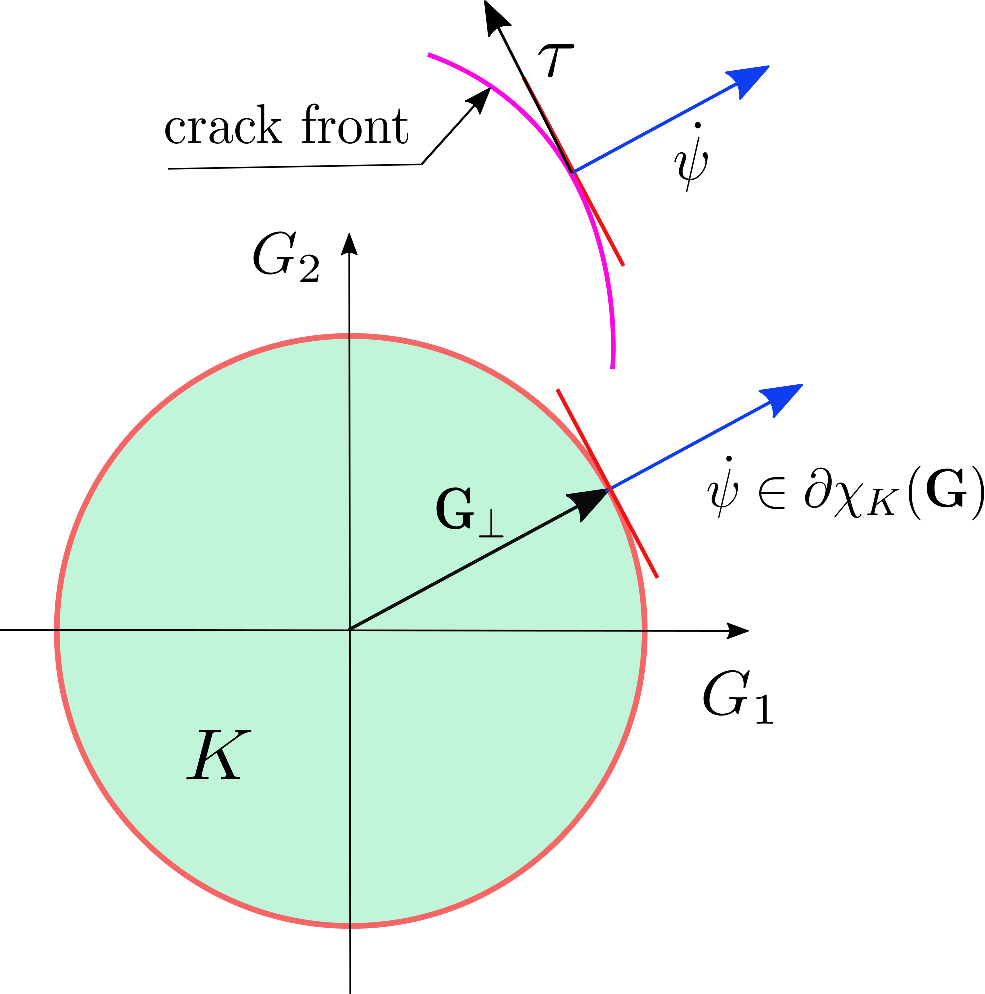}
	\caption{Normality law}
	\label{FigNormalityLaw}
\end{figure}

We call \textbf{driving force} the functional derivative 
$$  \bm{G}= - D_{\bm{\psi}} H 
$$
Denoting  $\bm{\tau}$ the unit tangent vector to the crack front, we introduce the deviatoric force  (Figure \ref{FigNormalityLaw})
$$\bm{G}_\perp = \bm{G} - (\bm{G} \cdot \bm{\tau}) \, \bm{\tau}
$$
the \textbf{crack stability domain}
$$ K = \left\lbrace \bm{G} \;\; \mbox{such that} \;\; 
     \parallel \bm{G}_\perp \parallel \leq G_c
       \right\rbrace 
$$
and the \textbf{crack extension rule} 
$$ \dot{\bm{\psi}} \in \partial \varphi (\bm{G}) 
     = \partial I_K (\bm{G})
$$

With this choice of constitutive law and withour friction on the crack sides, we particularize the variational SBEN principle (\ref{Functional of the SBEN principle}) in the form

\vspace{0.5cm}

\textbf{Symplectic BEN principle for cracks.} \textit{The natural evolution of the system minimizes}
\begin{equation}
\Pi(\bm{\xi}) = \int_{0}^{T} \left\{ 
   \int_{c_t} G_c \parallel \dot{\bm{\psi}} \parallel \, \mbox{d}s
- \left\langle \dot{\bm{\psi}}, \bm{G} \right\rangle
\right\} \mbox{ d} t
\label{SBEN 3 Pi (x) =}
\end{equation}
\textit{among the admissible curves $t \mapsto \bm{\xi} (t) = (\bm{u}, \bm{\psi}) $ such that} 
$ \nabla \cdot \bm{\sigma} +  \bm{f} = \rho \, \ddot{\bm{u}} $, $ \parallel \bm{G}_\perp \parallel \leq G_c$
\textit{and the minimum is zero.} 

\vspace{0.5cm}

\begin{figure}[ht!]
	\centering
	\includegraphics[scale=0.30]{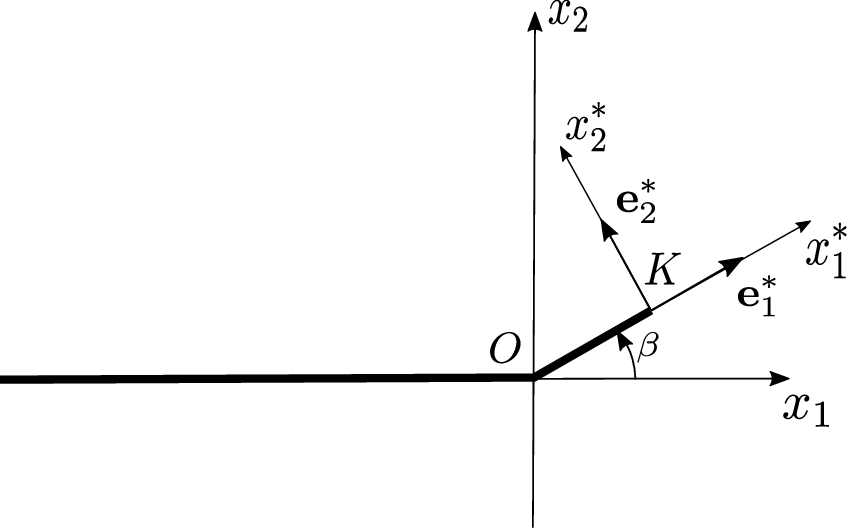}
	\caption{Kinked crack}
	\label{FigKinkImplicit}
\end{figure}

An important step is to check the validity of the normality law with respect to the experimental results on PMMA specimens \cite{Richard 1984}. A good test is to considered an initial straight crack in a plate which, under mixed mode loading, may extend suddenly in a direction deviating of an angle $\beta$ from the original one by kinking (Figure \ref{FigKinkImplicit}). The length of the kink crack is taken very small. The idea is to use the incremental formulation of the previous section with only one step. 

\begin{figure}[ht!]
	\centering
	\includegraphics[scale=.20]{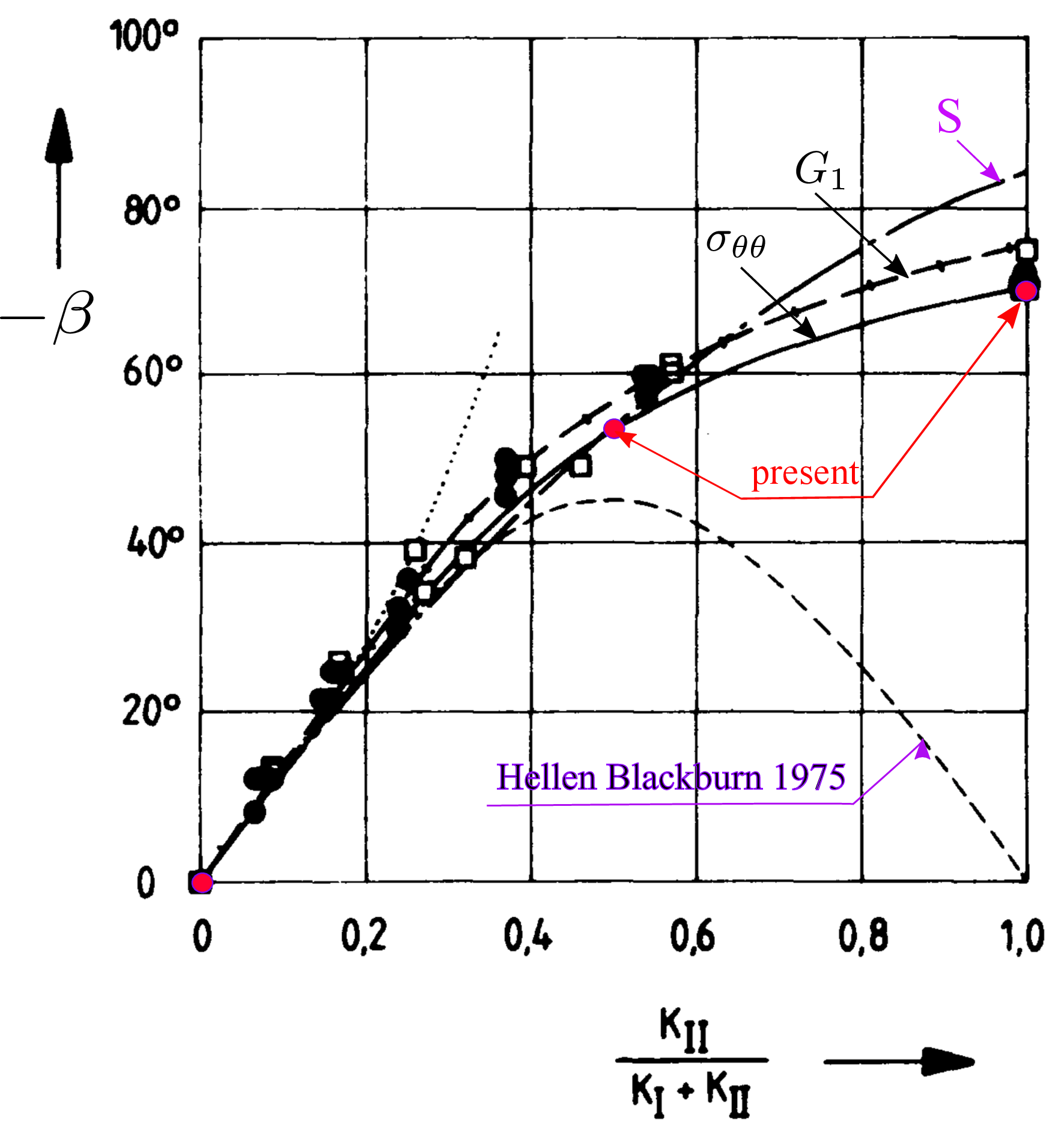}
	\caption{Comparison to the experimental results}
	\label{FigComparisonExperimentalResults}
\end{figure}

In the work of Hellen and Blackburn \cite{Hellen 1975}, Euler \textbf{explicit scheme} is used that can read with simplified notations
 $$    \Delta \bm{\psi} = \dot{\bm{\psi}} \, \Delta t 
       = \lambda \, \bm{G} \Delta t
 $$
As shown in \cite{de Saxce 2022}, the predictions are clearly disastrous (Figure \ref{FigComparisonExperimentalResults}). Our opinion is that the constitutive law must not be \textit{a priori} rejected but it is the explicit scheme which is problematic.

The problem of the kinked crack were studied by many authors. It is an awkward  problem of Elasticity and only approximated formula were proposed to express the Stress Intensity Factors (SIFs) $K^*_i$ at the kink crack tip in terms of the SIFs $K_i$ at the initial crack tip. At the limit of vanishing length of the kink crack, we adopt in the sequel the expression proposed in \cite{Cotterell 1980}. In \cite{de Saxce 2022}, we consider the \textbf{implicit scheme} 
 \begin{equation}
    \Delta \bm{\psi} = \dot{\bm{\psi}}^* \, \Delta t 
       = \lambda \, \bm{G}^* \Delta t
     \label{implicit scheme star}
 \end{equation}
In the frame of origin $K$ at the kink crack tip, the axis $Kx^*_1$ in the direction ahead the kink crack and the axis $Kx^*_2$ perpendicular to $Kx^*_1$ within the plane (Figure \ref{FigKinkImplicit}), the driven force is given in terms of stress intensity factors (SIFs) by
 $$G^*_1 =\frac{1}{\bar{E}} \, \left[(K^*_{I})^2 + (K^*_{II})^2\right] ,\qquad
   G^*_2 = -  \frac{1}{\bar{E}} \, 2\, K^*_{I} K^*_{II}
 $$
 where $ \bar{E} = 4 \, E / (1 + \nu) \, (1 + \kappa)$ with  $\kappa = 3 - 4 \, \nu$ in plane strain and $\kappa = (3 - \nu) / (1 + \nu)$ in plane stress. The normality law (\ref{implicit scheme star}) entails 
 \begin{equation}
  \bar{E}\, G^*_2 = - 2\, K^*_{I} K^*_{II} = 0, \qquad
 \bar{E}\, \parallel \bm{G}^* \parallel 
   =(K^*_{I})^2 + (K^*_{II})^2 ,\qquad
     \label{implicit scheme star 2}
 \end{equation}
  To satisfy the former condition, two scenarios may be considered for the crack extension:
 \begin{enumerate}
     \item \textbf{Scenario 1.} $\beta = \beta_1$, solution of $ K^*_{II} (\beta_1) =  0 $, and $\bar{E}\, \parallel \bm{G}^* \parallel =  (K^*_{I} (\beta_1))^2  $
     \item \textbf{Scenario 2.} $\beta = \beta_2$, solution of $ K^*_{I} (\beta_2) =  0 $, and $\bar{E}\, \parallel \bm{G}^* \parallel =  (K^*_{II} (\beta_2))^2  $     
 \end{enumerate}
 According to the stability criterion, the crack extends for the inclination $\beta_i$ with the maximum driven force magnitude. Then  scenario 1 is realized if $(K^*_{I} (\beta_1))^2 > (K^*_{II} (\beta_2))^2  $  and  scenario 2 otherwise. The details of the calculations are given in \cite{de Saxce 2022} and the present predictions are in very good agreement with the experiments as shown in Figure \ref{FigComparisonExperimentalResults}. 
 
 \begin{figure}[ht!]
	\centering
\includegraphics[scale=.30]{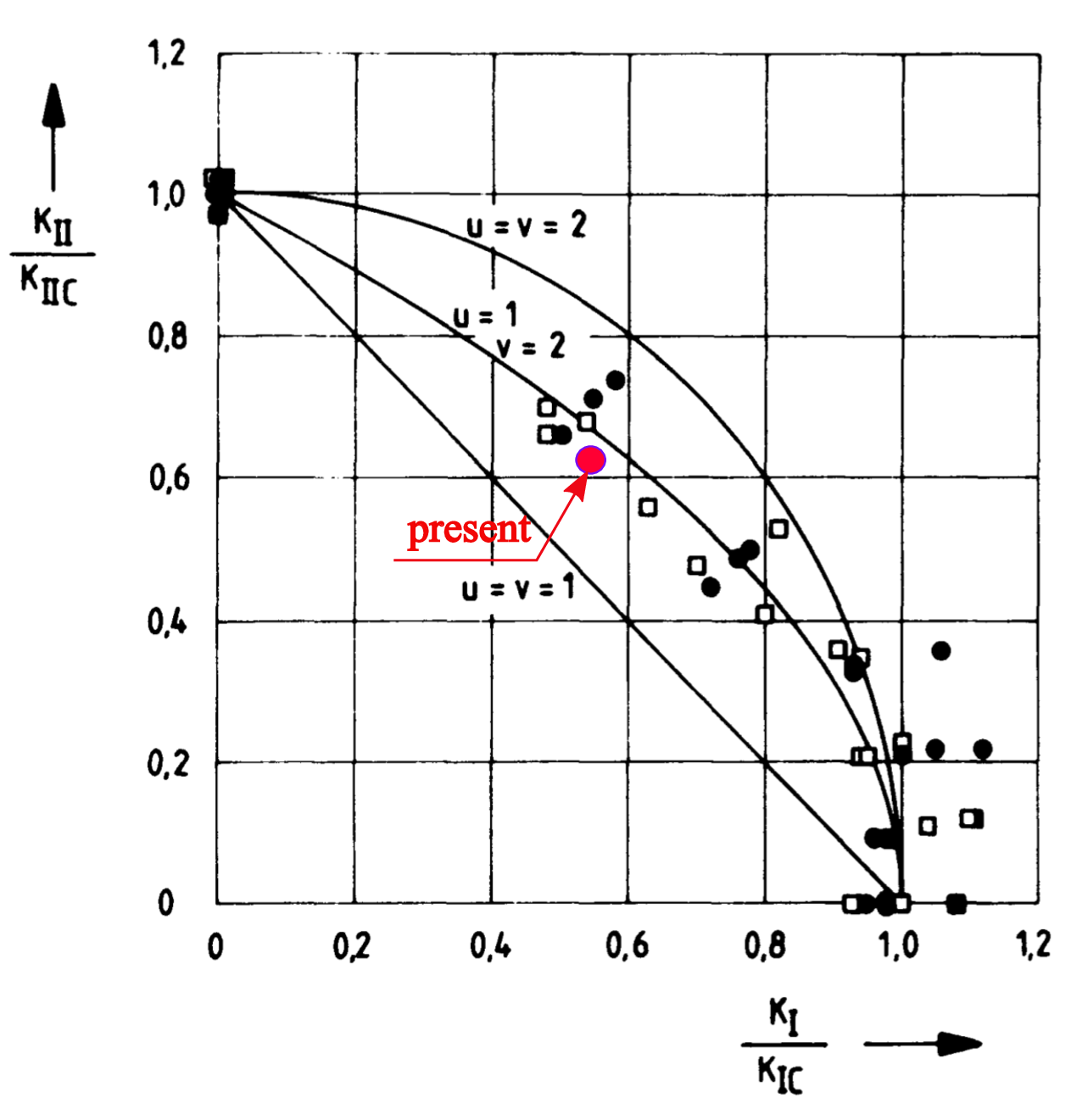}
	\caption{Richard empirical criterion }
	\label{FigRichardKi}
\end{figure}

 We conclude that:
 \begin{itemize}
\item On the basis of the experimental data, Richard proposed in \cite{Richard 1984} an empirical criterion in terms of the SIFs at the original crack tip (Figure \ref{FigRichardKi})
 $$ \frac{K_{I}}{K_{Ic}} + \left( \frac{K_{II}}{K_{IIc}} \right)^2 = 1
 $$
 According to (\ref{implicit scheme star 2}), the crack extends when
 $$ (K^*_{I})^2 + (K^*_{II})^2 = \bar{E}\, G_c
 $$
 that gives, for the scenario 1 of mode I, $K^2_{I} = \bar{E}\, G_c $ then
 $$ K^2_{Ic} = \bar{E}\, G_c
 $$
 and, for the scenario 1 of mode II,  $4 \, K^2_{II} / 3 = \bar{E}\, G_c = K^2_{Ic}$ then
 $$ \frac{K_{IIc}}{K_{Ic}} = \frac{\sqrt{3}}{2} = 0.86
 $$
 a value better than $\sqrt{2/3} = 0.81$ proposed in \cite{Fett 2002}, by comparison with the experimental data covering the range from $0.88$ to $0.95$. Moreover, our prediction for the mixed mode $K_{I} = K_{II}$ leads to values $\frac{K_{I}}{K_{Ic}} = 0.54 $ and $\frac{K_{II}}{K_{IIc}} = 0.62$ closed to the curve of Richard criterion.
\item At least in this example, the relevant scenario corresponds always to the \textbf{Principle of Local Symmetry} $ K^*_{II} (\beta) =  0 $  proposed by Goldstein and Sagalnik \cite{Goldstein 1974}. There can be no assurance that it is valid for arbitrary cracks. We recommend to replace it always by the normality law.
\item Our criterion should not be confused with the \textbf{Maximum Energy Release Rate criterion} for which the maximum must be searched among all the directions while with our criterion the maximum is found only among the directions satisfying the normality law, according to the different scenarios.
 \item The implicit scheme must be preferred to the explicit scheme in fracture mechanics.
\end{itemize}
 
Finally, we come back to the problem with possible friction between the crack sides. A crack being a material discontinuity, we must distinguish  two material surfaces $\Gamma^+_t$ and $\Gamma^-_t$ that occupy the same position as $\Gamma_t$ but are the two sides of the crack and have opposite unit normal vectors, exterior to $\Omega_t$: $\bm{n} = \bm{n}^- = - \bm{n}^+$. Let $\bm{u}^\pm$ (resp. $\bm{\sigma}^\pm $) be  the value of the displacement (resp. stress tensor) on $\Gamma^\pm_t$, $\left[\bm{u}\right] = \bm{u}^+ - \bm{u}^-$ be the relative displacement and $\bm{r} = \bm{\sigma}^+ \cdot \bm{n}$ be the reaction subjected to $\Gamma^+_t$ from $\Gamma^-_t$.  The unilateral contact with friction between the crack sides is governed by  the Coulomb's law of Section \ref{Section - Bipotentials}.

Applying the method of construction of the symplectic bipotential of  the previous section for the bipotential (\ref{bipo Coulomb}), we obtained in \cite{de Saxce 2026} the corresponding variational principle

\vspace{0.5cm}

\textbf{BEN principle for cracks with friction contact between the crack sides.} \textit{The natural evolution of the system minimizes}
\begin{equation}
\Pi(\bm{\xi}) = \int_{0}^{T} \left\{
   \int_{\Gamma_t} (b (- \left[ \dot{\bm{u}} \right] , \bm{r} ) + \left[ \dot{\bm{u}} \right] \cdot \bm{r} ) \, \mbox{d} \Gamma 
   + \int_{c_t} G_c \parallel \dot{\bm{\psi}} \parallel \, \mbox{d}s
- \left\langle \dot{\bm{\psi}}, \bm{G} \right\rangle
\right\} \mbox{ d} t
\label{SBEN 4 Pi (x) =}
\end{equation}
\textit{among the admissible curves $t \mapsto \bm{\xi} (t) = (\bm{u}, \bm{\psi}) $ such that} 
$ \nabla \cdot \bm{\sigma} +  \bm{f} = \rho \, \ddot{\bm{u}} $, $ \parallel \bm{G}_\perp \parallel \leq G_c$
\textit{and the minimum is zero.} 

\vspace{0.5cm}

 \section{Conclusions}
 
 \begin{itemize}
\item The BEN principle turns out to be a variational approach that is suited to a large spectrum of applications:
in smooth and non smooth mechanics, Lagrangian and Eulerian specification, for solids and fluids.
\vspace{0.2cm}
\item Our method to construct variational principles can be summarized in three words: \\
"derivating, integrating, minorizing",

"derivating" because we need a derivative of the Hamiltonian $H$ which must be variational because $H$ is a functional, symplectic if we consider the dynamics, in the sense of the jet theory if we use the Eulerian specification; 
"Integrating" to obtain the BEN functional;
"Minorizing" or majorizing if we are working with convex functionals. 
\vspace{0.2cm}
\item This approach is not simply a way to get weak formulations of the mechanics but it is also a source of ideas to give rise to some new research tracks on the modelization of constitutive laws
\end{itemize}

\vspace{0.5cm}

\textbf{Aknowledgments}

\vspace{0.3cm}

This work was performed thanks to the research project BIpotentiels G\'{e}n\'{e}ralis\'{e}s pour le principe variationnel de Brezis-Ekeland-Nayroles en m\'{e}canique (BigBen) supported by the Agence Nationale de la Recherche (Project ANR-22-CE51-0034-04). I thank Vincent Acary for the invitation, during a seminar on 5 February 2026 at the INRIA in Grenoble, to give a presentation that lies at the origin of this paper.

%%%%%%%%%% FRONT LINE ==========================
% TO DO (Figure Coulomb law) 

% TODO dans Slides INRIA Grenoble : #18 figure \lbrack dot{u} \rbrack
% TODO dans Slides INRIA Grenoble : #19 \lbrack dot{u} \rbrack

%%%%%%%%%%%%%%%%%%%%%%%%%%%%%%%%%%%%%%%%%%%%%%%%%%%

\end{document}